\documentclass[psfig,preprint,eqsecnum,preprintnumbers,nofootinbib,byrevtex,prd,aps,showpacs,showkeys,groupedaddress,floatfix]{revtex4}
\usepackage{bm}
\usepackage{graphics}
\usepackage{graphicx}
\usepackage{epsfig}
\usepackage{amssymb}
\usepackage{amsmath}

\newcommand\ba{\begin{eqnarray}}
\newcommand\ea{\end{eqnarray}}

\begin{document}

\date{}
\title{Bound states of the $D$-dimensional
Schr\"{o}dinger equation for the generalized Woods-Saxon potential}
\author{V.~H.~Badalov$^{1}$}\email{E-mail:badalovvatan@yahoo.com}
\author{B.~Baris$^{2}$}
\author{K.~Uzun$^{3}$}
\affiliation{$^{1}$Institute for Physical Problems, Baku State
University, Z. Khalilov st. 23, AZ-1148, Baku,
Azerbaijan\\
 $^{2}$\ Faculty of Arts and Sciences, Giresun University, 28100, Turkey \\
$^{3}$\ Department of Physics, Karadeniz Technical University,61080,
Trabson, Turkey}


\begin{abstract}
In this paper, the approximate analitical solutions of the
hyper-radial Schr\"{o}dinger equation are obtained for the
generalized Wood-Saxon potential by implementing the Pekeris
approximation to surmount the centrifugal term. The energy
eigenvalues and corresponding hyper-radial wave functions are found
for any angular momentum case via the Nikiforov-Uvarov (NU) and
Supersymmetric quantum mechanics (SUSY QM) methods. Hence, the same
expressions are obtained for the energy eigenvalues, and the
expression of hyper-radial wave functions transformed each other is
shown owing to these methods. Furthermore, a finite number energy
spectrum depending on the depths of the potential well $V_{0}$ and
$W$, the radial $n_{r}$ and $l$ orbital quantum numbers and
parameters $D,a,R_{0}$  are also identified in detail. Finally, the
bound state energies and the corresponding normalized hyper-radial
wave functions for the neutron system of the a $^{56} Fe$ nucleus
are calculated in $D=2$  and $D=3$, as well as the energy spectrum
expressions of other highest dimensions are identified by using the
energy spectrum of $D=2$ and $D=3$.
\end{abstract}

\pacs{03.65.Ge, 03.65.-W, 03.65.Fd, 02.30.Gp} \keywords{Bound
states, Generalized Woods-Saxon potential, Nikiforov - Uvarov
method, Supersymmetric Quantum Mechanics}

\maketitle

\section{\bf Introduction}
The Woods-Saxon potential was first proposed by R.D.Woods and
D.S.Saxon in 1954 to describe the elastic scattering of 20 MeV
protons on heavy nuclei ~\cite {Woods}. It is well known that the
Woods-Saxon potential which is considered in the present study is
one of the most realistic short-range potentials in physics. Since
then, the Woods-Saxon potential has attracted a great deal of
interest over the years and has been one of the most useful models
for determining the single particle energy levels of nuclei ~\cite
{Bohr, Gomez, Massen, Koura} and the nucleus-nucleus interactions
~\cite {Brandan, Satchler, Khoa, Boztosun, Kucuk}. Furthermore, the
exactly or quasi-exactly solutions of the Woods-Saxon type potential
for the wave equations (Schr\"{o}dinger, Dirac, Klein-Gordon) are of
high scientific importance in the conceptual understanding of the
interactions between the nucleon and the nucleus for both the bound
and resonant states. The modified version of the Woods-Saxon
potential consists of the volume (standard) Woods-Saxon and its
derivative called the Woods-Saxon surface potential and is given by
~\cite {Zaichenko, Fakhri, Berkdemir, Gonul, Bayrak1}:

\begin{equation}
V(r)=-\frac{V_{0} }{1+e^{\frac{r-R_{0} }{a} } }
+\frac{We^{\frac{r-R_{0} }{a} } }{\left(1+e^{\frac{r-R_{0} }{a} }
\right)^{2}},\, \, \, \, \, \, \, \, (a<<R_{0} ),
\end{equation}
where $V_{0}$  and $W$  represent the depths of the potential well,
$R_{0}$ and  $a$ are the radius of the potential and the width of
the surface diffuseness, respectively. The surface term in the
generalized Woods-Saxon potential induces an extra potential pocket
especially at the surface region of the potential, and this pocket
is so significant to explain the elastic scattering of some nuclear
reactions ~\cite {Brandan, Satchler, Khoa, Boztosun, Kucuk}.
Moreover, the Woods-Saxon surface potential induces a potential
barrier for $W<0$  so that it may be used in the explanation of the
resonant states (quasi-bound states) in nuclei. There are some
special cases of the generalized Woods-Saxon potential: the GWS
potential is reduced to the standard Woods-Saxon form ~\cite
{Flugge} for $W=0$  and the square well potential for $W=0$ and
$a\longrightarrow 0$ ~\cite {Flugge}. Furthermore the GWS potential
is reduced to the Rosen-Morse potential ~\cite {Rosen} for $R_{0}=0$
~\cite {Berkdemir}.

The exact analytical solutions of the wave equations such as
Schr\"{o}dinger, Dirac and Klein-Gordon with a physical potential is
of paramount importance in quantum physics since the wave function
and its associated eigenvalues contain all necessary information for
full description of a quantum system. It is well-known that the
exact analytical solutions of the this equations are very important,
since a bound states of the wave function is more convenient than
the wave function obtained by numerical calculation in explaining
the behavior of the system under consideration. Note that, along the
years, there was a huge amount of research effort to solve exactly
the radial Schr\"{o}dinger equation for all values of $n_{r}$  and
$l$ quantum numbers. Unfortunately, it could only be possible for a
very limited number of potentials such as harmonic oscillators,
Coulomb, Kratzer potentials and so on ~\cite {Flugge} with a
centrifugal term for $l\neq0$. In this way, there are several
established analytical methods, including Polynomial solution ~\cite
{Flugge, Landau, Greiner}, Nikiforov-Uvarov method (NU) ~\cite
{Nikiforov}, Supersymmetric quantum mechanics method (SUSY QM)
~\cite {Cooper, Morales}, and Asymptotic iteration method (AIM)
~\cite {Ciftci1,Ciftci2,Ciftci3,Bayrak2,Bayrak3,Bayrak4,Ciftci4}, in
order to solve analytically the radial Schr\"{o}dinger equation for
$l\neq0$ within various potentials. G.Levai et al. suggested a
simple method for the proposed potentials for which the
Schr\"{o}dinger equation can be solved exactly with special
functions ~\cite {Levai} and presented relationship between the
introduced formalism and supersymmetric quantum mechanics ~\cite
{Cooper}. Furthermore, in order to solve the Schr\"{o}dinger
equation applicable to problems of non-perturbative nature, P.Amore
et al. introduced a novel method ~\cite {Amore}. Thus, this method
was applied to calculate the energies and wave functions of the
ground and first excited state of the quantum anharmonic potential.
Although S.Fl\"{u}gge gave an exact expression for the wave function
and suggested a graphical method for the energy eigenvalues at
$l=0$, this potential could not be solved exactly without using any
approximation for $l\neq0$ yet ~\cite {Flugge}.

In recent years, the Nikiforov-Uvarov (NU)~\cite {Nikiforov} and
Supersymmetric quantum mechanics (SUSY QM) ~\cite{Cooper,Morales}
methods with various approximations have been proposed for solving
the Schr\"{o}dinger type equations analytically. Many papers show
the power and simplicity of both of these methods in solving central
and noncentral potentials ~\cite
{Badalov1,Badalov2,Badalov3,Badalov4,Ahmadov1,Ahmadov2,Ahmadov3,Ahmadov4,Ikhdair1,Ikhdair2,Ikhdair3,Ikhdair4}.
Hence, NU method allows one to solve the second order linear
differential equation by reducing to a generalized equation of
hypergeometric type which is a second order type homogeneous
differential equation with polynomials coefficients of degree not
exceeding the corresponding order of differentiation, and SUSY QM
method is based on determining the eigenstates of known analytically
solvable potentials using algebraic operator formalism without ever
having to solve the Schr\"{o}dinger differential equation by
standard series technique. That is why, it would be more significant
to solve the non-relativistic radial Schr\"{o}dinger equation with
using both of these methods for the generalized Woods-Saxon
potential for $l\ne 0$, since it has been extensively utilized to
describe the bound and continuum states of the interacting systems.
In this way, one can need to obtain the energy eigenvalues and
corresponding eigenfunctions of the one particle problem within this
potential. The NU method was used in ~\cite {Berkdemir} to solve the
radial Schr\"{o}dinger equation for the generalized Woods-Saxon
potential for $l=0$. In this paper, it has been noted that there are
inconsistent between the analytical and the numerical results for
the GWS potential with the $l=0$ state ~\cite {Berkdemir}. Hence, in
mentioned study, the authors made errors in application of the NU
method, and these errors have led to incorrect results ~\cite
{Editorial}. Later authors made also similar errors in application
of the Nikiforov-Uvarov and Asymptotic iteration methods in the
following works ~\cite {Ikhdair5,Arda1,Arda2, Ikhdair6}.

Note that, there are some efforts to obtain the approximate
analytical solutions of the wave equations in terms of the $l\neq0$
case; the most widely used approximation is introduced by Pekeris
~\cite {Pekeris} for the exponential type potential so that this
approximation is based on the expansion of the centrifugal barrier
in a series of exponentials depending on the internuclear distance
up to second order. Another approximation was proposed for the
centrifugal term by Greene and Aldrich ~\cite {Greene}

\begin{equation}
\frac{1}{r^{2}}=\delta^{2}\frac{e^{\delta r}}{\left(1-e^{\delta
r}\right)^{2}},
\end{equation}
However, this approximation is only effective for small values of
the screening parameter $\delta$ ~\cite {Bayrak4}.

In our previous works ~\cite {Badalov1,Badalov2,Badalov3}, in the
first time we have given the comprehensive information of how to
obtain analytically the exact energy eigenvalues and the
corresponding wave functions of the radial Schr\"{o}dinger and the
radial Klein-Gordon equations with Woods-Saxon potential via NU
method. Furthermore, the hyper-radial Schr\"{o}dinger equation with
Woods-Saxon potential was solved within the context of the NU, AIM,
and SUSY QM methods through the Pekeris approximation to the
centrifugal potential, and the energy eigenvalues and corresponding
radial wave functions are found for any arbitrary $l$ state as well
~\cite {Badalov4}. In these works ~\cite
{Badalov1,Badalov2,Badalov3,Badalov4}, the below approximation
scheme - the Pekeris approximation ~\cite {Pekeris} - was proposed
for $V_{l}(r)=\frac{\hbar^{2} \tilde{l}(\tilde{l}+1)}{2\mu r^{2} }$
the centrifugal potential in any arbitrary $l$ state:

\begin{equation}
\frac{1}{r^{2} } =\frac{1}{R_{0}^{2} } \left(C_{0} +\frac{C_{1}
}{1+e^{\frac{r-R_{0} }{a} } } +\frac{C_{2}
}{\left(1+e^{\frac{r-R_{0} }{a} } \right)^{2} } \right)\, ,
\end{equation}
where $C_{0}, C_{1}, C_{2}$  quantities dependent on $R_{0}, a$
specific potential parameters were defined by comparing both sides
of Eq.(1.1) expression expanding in the Taylor series around the
point $r=R_{0}$. But, in Ref.~\cite {Badalov5} using NU and SUSY QM
methods to solve successfully the $D$ -dimensional Schr\"{o}dinger
equation with Woods-Saxon potential by applying the Pekeris
approximation to the centrifugal potential $V_{l}(r)$ based on the
settings $C_{0}, C_{1}, C_{2}$ quantities which are determined
around the point $r=r_{min}$  of the effective potential
$V_{eff}(r)$ in any arbitrary $l$ state. There it became clear that
the system has not the bound states for $n_{r}=0$, $l=0$  when $D=3$
and $D=4$, and its reason is related with the standard Woods-Saxon
potential cannot describe the system fully. As a way out of this,
one need to solve this issue with utilizing the modified version of
the standard Woods-Saxon potential such as the generalized
Woods-Saxon potential, or the spin and pseudospin symmetries in the
standard Woods-Saxon potential ~\cite
{Hecht,Arima,Ginocchio1,Zhou,Ginocchio2,Liang}.

In present paper, our aim is to solve of the hyper radial
Schr\"{o}dinger  equation for the generalized Woods-Saxon potential
and to obtain the eigenvalues of energy and corresponding
eigenfunctions for arbitrary $l$ angular momentum, by using the
Nikiforov-Uvarov and Supersymmetric quantum mechanics methods.

The remainder of the paper is organized as follows: After this
introduction, the detailed description of the  $D$-dimensional SE
with the generalized Woods-Saxon potential is given in Section II.
Next, the implementation of Nikiforov-Uvarov (NU) and Supersymmetric
quantum mechanics (SUSY QM) methods to  hyper-radial SE which will
be a useful guideline for investigators are presented in Section III
and IV, respectively. Then in Section V, the results and discussion
are presented. Finally, the paper is concluded with brief summary in
Section VI.

\section{\bf  The $D$-dimensional Schr\"{o}dinger  equation with the Woods-Saxon potentials}

In the $D$-dimensional $D\geq 2$ space, the Schr\"{o}dinger equation
with spherically symmetric potential $V(r)$ is of the form ~\cite
{Avery}

\begin{equation}
\left(-\frac{\hbar ^{2} }{2\mu } \nabla _{D}^{2} +V(r)-E_{n_{r}l}
\right)\psi _{n_{r}lm} (r,\Omega _{D} )=0\, ,
\end{equation}
where $D$ is dimension of space, $\mu $ is the reduced mass, $r$ is
hyperradius, $\Omega _{D}=(\theta _{1} \, ,\, \theta _{2} \, ,\,
\ldots ,\, \theta _{D-2} \, ,\, \phi)$ is hyperangular, $\hbar \;$
is the Planck's constant, $\triangle_{D}\equiv\nabla^{2}_{D}$ is the
Laplasian operator and

\begin{equation}
\psi _{n_{r}lm} (r,\Omega _{D} )=R_{n_{r}l}(r)Y_{lm} (\Omega _{D}
)\, .
\end{equation}
The Laplasian operator divides into a hyper-radial part $r^{1-D}
\frac{\partial }{\partial r} \left(r^{D-1} \frac{\partial }{\partial
r} \right)$ and an angular part $-\frac{\hat{L}_{D}^{2} }{\hbar ^{2}
r^{2} } $, i.e.

\begin{equation}
\nabla _{D}^{2} =r^{1-D} \frac{\partial }{\partial r} \left(r^{D-1}
\frac{\partial }{\partial r} \right)-\frac{\hat{L}_{D}^{2} }{\hbar
^{2} r^{2} },
\end{equation}
where $\hat{L}_{D} $ is the grand orbital angular momentum operator.
The eigenfunctions of $\hat{L}_{D}^{2} $ are the hyper-spherical
harmonics

\begin{equation}
\hat{L}_{D}^{2} Y_{lm} (\Omega _{D} )\, =\hbar ^{2} l(l+D-2)Y_{lm}
(\Omega _{D} )\, ,
\end{equation}
where $l$ is the angular momentum quantum number.

After substituting the Eqs.(2.2) - (2.4) into (2.2) and using $\psi
_{n_{r}lm} (r,\Omega _{D} )$ as the eigenfunction of
$\hat{L}_{D}^{2} $ with eigenvalue $\hbar ^{2} l(l+D-2)$, we obtain
an equation known as the hyper-radial Schr\"{o}dinger equation with
generalized Woods-Saxon potential

\begin{equation}
\frac{d^{2} R_{n_{r}l} (r)}{dr^{2} } +\frac{D-1}{r}
\frac{dR_{n_{r}l} (r)}{dr} +\frac{2\mu }{\hbar ^{2} } \left[E_{nl}
-V(r)\, \, -\frac{\hbar ^{2} l(l+D-2)}{2\mu r^{2} }
\right]R_{n_{r}l} (r)=0\, ,\, \, (0\le r<\infty ).
\end{equation}
Introducing a new function $u_{n_{r}l} (r)=r^{\, \frac{D-1}{2} }
R_{n_{r}l} (r)$ and a new parametr $\tilde{l}=l+\frac{D-3}{2}$ ,
Eq.(2.5) reduces to

\begin{equation}
\frac{d^{2} u_{n_{r}l} (r)}{dr^{2} } +\frac{2\mu }{\hbar ^{2} }
\left[E_{n_{r}l} -V_{eff} (r)\right]u_{n_{r}l} (r)=0,
\end{equation}
where $V_{eff} (r)$ is effective potential, i.e.

\begin{equation}
V_{eff}(r)=V(r)+\frac{\hbar ^{2} \tilde{l}(\tilde{l}+1)}{2\mu r^{2}
}   .
\end{equation}

Equation (2.7) has the same form as the equation for a particle in
one dimension, except for two important differences. First, there is
a repulsive effective potential proportional to the eigenvalue of
$\hbar ^{2} \tilde{l}(\tilde{l}+1)$. Second, the radial function
must satisfy the boundary conditions $u(0)=0$ and $u(\infty )=0$.

It is well-known that the hyper-radial Schr\"{o}dinger equation
cannot be solved exactly for this potential at the value $l\ne 0$ by
using the standard methods as SUSY QM, AIM, NU and so on. From
Eq.(2.7), it is seen that the effective potential is combination of
the exponential and inverse square potentials which cannot be solved
analytically. That is why, the Pekeris approximation which is the
most widely used and convenient for our purposes can be taken in
order to solve this issue. This approximation is based on the
expansion the series for exponential cases depending on the
internuclear distance of the centrifugal barrier, and there the
terms up to second order are considering.

After introducing new variable  $x=\frac{r-R_{0} }{R_{0} }$  or
$r=R_{0} (1+x)$, the generalized Woods-Saxon potential in that case
has the following form:

\begin{equation}
V(r)=-\frac{V_{0} }{1+e^{\alpha x}}+\frac{We^{\alpha
x}}{(1+e^{\alpha x})^{2}},
\end{equation}
where $\alpha=\frac{R_{0} }{a}$. The extreme point of the
generalized Woods-Saxon potential $V(r)$  is defined as

\begin{equation}
exp\left(\alpha x\right) =\frac{W-V_{0}}{W+V_{0}}
\quad\quad\quad\quad  or \quad\quad\quad exp \left(
\frac{r-R_{0}}{a}\right )= \frac{W-V_{0}}{W+V_{0}}.
\end{equation}
and their solution of Eq.(2.9) is the following form:

\begin{equation}
x=x_{min}=\frac{1}{\alpha}\ln\left(\frac{W-V_{0}}{W+V_{0}}\right)
\quad\quad or \quad\quad
r=r_{min}=R_{0}+a\ln\left(\frac{W-V_{0}}{W+V_{0}}\right).
\end{equation}

It must be noted, there the conditions $W>V_{0}$ and $W<-V_{0}$  is
obtained based on the relation $exp\left(\alpha x\right)
=\frac{W-V_{0}}{W+V_{0}}$ because $exp\left(\alpha x\right)>0$  in
an arbitrary value of $x$ . If this condition is not satisfied,
there are not minimum point in the generalized Woods-Saxon potential
$V(r)$.

Let us expand centrifugal potential $V_{l} (r)$ in Taylor series
around the point of $x=x_{min}=x_{e} $ \;($r=r_{min}=r_{e}$):

\begin{equation}
\begin{array}{l} {V_{l} (r)=\frac{\hbar ^{2} \tilde{l}(\tilde{l}+1)}{2\mu r^{2}
} =\frac{\hbar ^{2} \tilde{l}(\tilde{l}+1)}{2\mu R_{0}^{2} } \cdot
\frac{\tilde{\delta }}{(1+x)^{2} } =\tilde{\delta
}\left[\frac{1}{(1+x_{l} )^{2} } -\frac{2}{(1+x_{l} )^{3} } \cdot
(x-x_{l} )+\right. } \\ {\left. \, \, \, \, \, \, \, \, \, \, \, \,
\, \, \, \, \, \, \, \, \, \, \, \, \, \, \, \, \, \, \, \, \, \, \,
\, \, \, \, \, \, \, \, \, \, \, \, \, \, \, \, \, \, \, \, \, \, \,
\, \, \, \, \, \, \, \, \, \, \, +\, \frac{3}{(1+x_{l} )^{3} } \cdot
(x-x_{l} )^{2} +o((x-x_{l} )^{3} )\right]} \end{array},
\end{equation}
where $\tilde{\delta }=\frac{\hbar ^{2} \tilde{l}(\tilde{l}+1)}{2\mu
R_{0}^{2} }$. By using the Pekeris approximation, $V_{l}(r)$ takes
the form ~\cite {Badalov1,Badalov2,Badalov3,Badalov4,Badalov5}

\begin{equation}
\tilde{V}_{l} (r)=\tilde{\delta }\left(C_{0} +\frac{C_{1}
}{1+e^{\alpha x} } +\frac{C_{2} }{(1+e^{\alpha x} )^{2} } \right).
\end{equation}
Let us expand the potential $\tilde{V}_{l} (r)$ in the Taylor series
around the point of $x=x_{min}=x_{e} $ \;($r=r_{min}=r_{e}$):

\begin{equation}
\begin{array}{l} {\tilde{V}_{l} (x)=\tilde{\delta }\left[C_{0} +\frac{C_{1} }{1+e^{\alpha x_{e} }
} +\frac{C_{2} }{(1+e^{\alpha x_{e} } )^{2} } -\left(\frac{\alpha
C_{1} \, e^{\alpha x_{e} } }{(1+e^{\alpha x_{e} } )^{2} }
+\frac{2\alpha C_{2} \, e^{\alpha x_{e} } }{(1+e^{\alpha x_{e} }
)^{3} } \right)(x-x_{e} )+ \right. } \\ \left.{\, \, \, \, \, \, \,
\, \, \, \, \, -\left(\frac{\alpha ^{2} C_{1} \, e^{\alpha x_{e} }
(1-e^{\alpha x_{e} } )}{2(1+e^{\alpha x_{e} } )^{3} } +\frac{\alpha
^{2} C_{2} \, e^{\alpha x_{e} } (1-2e^{\alpha x_{e} }
)}{(1+e^{\alpha x_{e} } )^{4} } \right)(x-x_{e} )^{2} +o((x-x_{e}
)^{3} )} \right]
\end{array}.
\end{equation}

In order to define the constants $C_{0} \, ,\, C_{1} $ and $C_{2}$,
we should  compare the compatible degrees of same order of $x$ in
Eqs.(2.11) and (2.13), and we get the following algebraic set of
equations:

\begin{equation}
\left\{\begin{array}{l} {C_{0} +\frac{C_{1} }{1+e^{\alpha x_{e} }
}+\frac{C_{2} }{(1+e^{\alpha x_{e} } )^{2} } =\frac{1}{(1+x_{e}
)^{2} } } \\ {\frac{\alpha C_{1} e^{\alpha x_{e} } }{(1+e^{\alpha
x_{e} } )^{2} } +\frac{2\alpha C_{2} e^{\alpha x_{e} }
}{(1+e^{\alpha x_{e} } )^{3} } =\frac{2}{(1+x_{e} )^{3} } } \\
{\frac{\alpha ^{2} C_{1} \, e^{\alpha x_{e} } (1-e^{\alpha x_{e} }
)}{2(1+e^{\alpha x_{e} } )^{3} } +\frac{\alpha ^{2} C_{2} \,
e^{\alpha x_{e} } (1-2e^{\alpha x_{e} } )}{(1+e^{\alpha x_{e} }
)^{4} } =-\frac{3}{(1+x_{e} )^{4} } } \end{array}\right. .
\end{equation}

From the Eq.(2.14), for $C_{0} \, ,\, C_{1} $ and $C_{2}$ constants
is obtained as:
\begin{equation}
\left\{\begin{array}{l} {C_{0} =\frac{1}{(1+x_{e} )^{2} }
+\frac{(1+e^{\alpha x_{e} } )^{2} }{\alpha e^{\alpha x_{e} }(1+x_{e}
)^{3} } \left[\frac{e^{-\alpha x_{e} } -3}{1+e^{\alpha x_{e} } }
+\frac{3e^{-\alpha x_{e} } }{\alpha (1+x_{e} )} \right]}
\\ {C_{1} =\frac{2(1+e^{\alpha x_{e} } )^{2} }{\alpha e^{\alpha
x_{e} } (1+x_{l} )^{3} } \left[2-e^{-\alpha x_{e} }
-\frac{3(1+e^{-\alpha x_{e} } )}{\alpha (1+x_{e} )} \right]} \\
{C_{2} =\frac{(1+e^{\alpha x_{e} } )^{3} }{\alpha e^{\alpha x_{e} }
(1+x_{e} )^{3} } \left[e^{-\alpha x_{e} } -1+\frac{3(1+e^{-\alpha
x_{e} } )}{\alpha (1+x_{e} )} \right]} \end{array}\right. .
\end{equation}

After Pekeris approximation, the effective potential takes as
following form:

\begin{equation}
\tilde{V}_{eff} (r)= V_{WS}(r)+\tilde{V}_{l}(r)= \tilde{\delta
}C_{0} -\frac{V_{0}+W-\tilde{\delta }C_{1} }{1+e^{\frac{r-R_{0} }{a}
} }+\frac{W+\tilde{\delta }C_{2} }{\left(1+e^{\frac{r-R_{0} }{a} }
\right)^{2} } .
\end{equation}

If we consider $x_{e} =0$ in Eq.(2.15) relations, the constants
$C_{0} \, ,\, C_{1} $ and $C_{2}$ becomes in the following form
~\cite {Badalov1,Badalov2,Badalov3,Badalov4}:

\begin{equation}
C_{0} \, =1-\frac{4}{\alpha } +\frac{12}{\alpha ^{2}},
 C_{1} \, =\frac{8}{\alpha }-\frac{48}{\alpha ^{2} }, C_{2} \, =\frac{48}{\alpha ^{2} }.
\end{equation}

Instead of solving the hyper-radial Schr\"{o}dinger equation for the
effective generalized Woods-Saxon potential $V_{eff}(r)$ given by
Eq.(2.7), we now solve the radial Schr\"{o}dinger equation for the
new effective potential $\tilde{V}_{eff}(r)$ given by Eq.(2.16)
obtained using the Pekeris approximation. Having inserted this new
effective potential into Eq.(2.6), we obtain

\begin{equation}
\frac{d^{2} u_{n_{r}l} (r)}{dr^{2} } +\frac{2\mu }{\hbar ^{2} }
\left[E_{n_{r}l} -\tilde{\delta }C_{0} +\frac{V_{0}+W-\tilde{\delta
}C_{1} }{1+e^{\frac{r-R_{0} }{a} } }-\frac{W+\tilde{\delta }C_{2}
}{\left(1+e^{\frac{r-R_{0} }{a} } \right)^{2} } \right]u_{n_{r}l}
(r)=0.
\end{equation}

If the Eq.(2.18) is rewrite by using a new variable of the form
$z=\left(1+e^{\frac{r-R_{0} }{a} } \right)^{-1} $, it will be
obtained as

\begin{equation}
z^{2}(1-z)^{2} u''(z)+z(1-z)(1-2z)u'(z)+\frac{2\mu a^{2}}{\hbar^{2}
}\left[E-\tilde{\delta }C_{0} +(V_{0}+W-\tilde{\delta }C_{1})
z-(W+\tilde{\delta }C_{2}) z^{2}\right]u(z)=0.
\end{equation}

If we use the following dimensionless notations

\begin{equation}
\varepsilon ^{2} =-\frac{2\mu \, a^{2} (E-\tilde{\delta}C_{0}
)}{\hbar ^{2} }
>0,\, \, \beta ^{2}=\frac{2\mu \, a^{2} (V_{0}+W-\tilde{\delta }C_{1}) }{\hbar ^{2} } >0,\,
\, \gamma ^{2} =\frac{2\mu \, a^{2} (W+\tilde{\delta }C_{2}) }{\hbar
^{2} } >0,
\end{equation}
we will get

\begin{equation}
u''(z)+\frac{1-2z}{z(1-z)} u'(z)+\frac{-\varepsilon ^{2} +\beta ^{2}
z-\gamma ^{2} z^{2}}{\left(z(1-z)\right)^{2} } u(z)=0\, \, ,\, \, \,
(0\le z\le 1)
\end{equation}
with real $\varepsilon >0$ \,($E<0$) for bound states; $\beta $ and
$\gamma $ are real and positive.

\section{\bf Solution of the hyper-radial Schr\"{o}dinger equation by Nikiforov-Uvarov Method}

According to the NU-method ~\cite {Nikiforov}, the Eq.(2.21) is
compared with the equation of $\psi ''(z)+\frac{\tilde{\tau
}(z)}{\sigma (z)} \psi'(z)+\frac{\tilde{\sigma }(z)}{\sigma ^{2}(z)}
\psi(z)=0,$ and the corresponding polinomials are obtained

\begin{equation}
\tilde{\tau }(z)=1-2z;\, \, \sigma (z)=z(1-z);\, \, \tilde{\sigma
}(z)=-\varepsilon ^{2} +\beta ^{2} z-\gamma ^{2} z^{2}    .
\end{equation}
The new function $\pi (z)=\frac{\sigma '(z)-\tilde{\tau (z)}}{2} \pm
\sqrt{\left(\frac{\sigma '(z)-\tilde{\tau (z) }}{2} \right)^{2}
-\tilde{\sigma (z)}+k\sigma (z)} $ in Ref. ~\cite {Nikiforov} can be
found by substituting Eq.(3.1) and taking $\sigma '(z)=1-2z$. Hence,
the function $\pi (z)$ is

\begin{equation}
\pi (z)=\pm \sqrt{\varepsilon ^{2} + (k-\beta ^{2})z-(k-\gamma ^{2}
)z^{2}} .
\end{equation}

Further, the constant parameter $k$ can be found by utilizing the
condition that the expression under the square root has a double
zero, i.e., its discriminant is equal to zero. Therefore, the
parameter $k$ can be obtained as

\begin{equation}
k=\beta ^{2} -2\varepsilon ^{2} \pm 2\varepsilon \sqrt{\varepsilon
^{2} -\beta ^{2} +\gamma ^{2} } .
\end{equation}

So, we arrive at the following four possible functions of $\pi (z)$

\begin{equation}
\pi (z)=\left\{\begin{array}{l} {\left(\varepsilon
-\sqrt{\varepsilon ^{2} -\beta ^{2} +\gamma ^{2} }
\right)z-\varepsilon \, \, , \, \, \, \, \, \, \, \, \, {\rm for}\,
\, \, k=\beta ^{2} -2\varepsilon ^{2} +2\varepsilon
\sqrt{\varepsilon ^{2} -\beta ^{2} +\gamma ^{2} } \, ,} \\
{-\left(\varepsilon +\sqrt{\varepsilon ^{2} -\beta ^{2} +\gamma ^{2}
} \right)z+\varepsilon \, \, ,\, \, \, {\rm for}\, \, \, k=\beta
^{2} -2\varepsilon ^{2} +2\varepsilon
\sqrt{\varepsilon ^{2} -\beta ^{2} +\gamma ^{2} } \, ,} \\
{\left(\varepsilon +\sqrt{\varepsilon ^{2} -\beta ^{2} +\gamma ^{2}
} \right)z-\varepsilon \, \, , \, \, \, \, \, \, \, \, \, {\rm
for}\, \, \, k=\beta ^{2} -2\varepsilon ^{2} -2\varepsilon
\sqrt{\varepsilon ^{2} -\beta ^{2} +\gamma ^{2} } \, ,} \\
{-\left(\varepsilon +\sqrt{\varepsilon ^{2} -\beta ^{2} +\gamma ^{2}
} \right)z+\varepsilon \, \, ,\, \, \, {\rm for}\, \, \, k=\beta
^{2} -2\varepsilon ^{2} -2\varepsilon \sqrt{\varepsilon ^{2} -\beta
^{2} +\gamma ^{2} } \, .}
\end{array}\right.
\end{equation}

Further, from the four possible forms of the function $\pi(z)$ we
select the one for which the function $\tau (z)=\tilde{\tau }(z) +
2\pi (z)$ has the negative derivative $\tau '(z)<0$ ~\cite
{Nikiforov} and root lies in the interval $(0,\, \, 1)$. Hence, we
can select the parameter $k$ and the function $\pi(z)$ to be

\begin{equation}
k=\beta ^{2} -2\varepsilon ^{2} -2\varepsilon \sqrt{\varepsilon ^{2}
-\beta ^{2} +\gamma ^{2} }
\end{equation}
and
\begin{equation}
\pi (z)=\varepsilon -\left(\varepsilon +\sqrt{\varepsilon ^{2}
-\beta ^{2} +\gamma ^{2} } \right)z \, ,
\end{equation}
which give the physicaly valid expression for the function  $\tau
(z)$:

\begin{equation}
\tau (z)=1+2\varepsilon -2\left(1+\varepsilon +\sqrt{\varepsilon
^{2} -\beta ^{2} +\gamma ^{2} } \right)z\,,\,\, \tau'
(z)=-2\left(1+\varepsilon +\sqrt{\varepsilon ^{2} -\beta ^{2}
+\gamma ^{2} } \right)<0.
\end{equation}

Then, by using the relations ~\cite {Nikiforov} $\lambda =k+\pi '(z)
\,$ and $\lambda =\lambda _{n} =-n\tau '-\frac{n(n-1)}{2} \sigma
''\, \, ,\, \, \, (n=0,\, 1,\, 2,\, ...\, )\,$ the following
expressions for the constant $\lambda$ are obtained, respectively:
\begin{equation}
\lambda =\beta ^{2} -2\varepsilon ^{2} -2\varepsilon
\sqrt{\varepsilon ^{2} -\beta ^{2} +\gamma ^{2} }-\varepsilon
-\sqrt{\varepsilon ^{2} -\beta ^{2} +\gamma ^{2} } ,
\end{equation}
\begin{equation}
\lambda =\lambda _{n_{r} } =2\left(\varepsilon +\sqrt{\varepsilon
^{2} -\beta ^{2} +\gamma ^{2} } \right)\, n_{r} +n_{r} (n_{r} +1).
\end{equation}
After comparing the Eq.(3.8) with Eq.(3.9)

$\beta ^{2} -2\varepsilon ^{2} -2\varepsilon \sqrt{\varepsilon ^{2}
-\beta ^{2} +\gamma ^{2} } -\varepsilon -\sqrt{\varepsilon ^{2}
-\beta ^{2} +\gamma ^{2} } =2\left(\varepsilon +\sqrt{\varepsilon
^{2} -\beta ^{2} +\gamma ^{2} } \right)\, n_{r} +n_{r} (n_{r} +1)$
we get

\begin{equation}
\varepsilon +\sqrt{\varepsilon ^{2} -\beta ^{2} +\gamma ^{2} }
+n_{r} +\frac{1}{2} -\frac{\sqrt{1+4\gamma ^{2} } }{2} =0
\end{equation}
or
\begin{equation}
\varepsilon +\sqrt{\varepsilon ^{2} -\beta ^{2} +\gamma ^{2} }=n'.
\end{equation}
Here

\begin{equation}
n'=-n_{r} +\frac{\sqrt{1+4\gamma ^{2} } -1}{2},
\end{equation}
$n_{r}$ is the radial quantum number $(n_{r} =0,\, 1,\, 2,\, ...)$.
From Eq.(3.11), one can find

\begin{equation}
\varepsilon =\frac{1}{2} \left(n'+\frac{\beta ^{2} -\gamma ^{2}
}{n'} \right).
\end{equation}

From the bound states $E<0$ and finite wave function, we obtain
$\varepsilon >0$ and $\varepsilon ^{2} -\beta ^{2} +\gamma ^{2}>0$,
i.e. $n'>0$, $n'-\varepsilon >0$ and $|\beta ^{2} -\gamma ^{2}|
<n'^{2} $. Hence, using Eqs.(3.12) and (2.20) there relations can be
replaced into the form:

\begin{equation}
0\le n_{r} <\frac{1}{2} \left(\sqrt{1+\frac{8\mu \, a^{2}W}{\hbar
^{2} }+\frac{4\tilde{l}(\tilde{l}+1)a^{2}C_{2} }{R_{0}^{2}}}
-1\right),
\end{equation}

\begin{equation}
\left|V_{0}-\frac{\hbar^{2}\tilde{l}(\tilde{l}+1)}{2\mu
R_{0}^{2}}\left(C_{1}+C_{2} \right)\right|<\frac{\hbar ^{2}}{2\mu
a^{2}}\left[\frac{1}{2} \left(\sqrt{1+\frac{8\mu a^{2}W}{\hbar^{2}
}+\frac{4\tilde{l}(\tilde{l}+1)a^{2}C_{2} }{R_{0}^{2}}}
-1\right)-n_{r}\right]^{2} ,
\end{equation}

\begin{equation}
V_{0} R_{0}^{2}\geq \frac{\hbar ^{2}
\tilde{l}(\tilde{l}+1)}{2\mu}\left(C_{0}+C_{1}+C_{2} \right) .
\end{equation}
Upon substituting with the values of $\varepsilon ,\, \, \beta ,\,
\gamma$,\, $n'$ and $\tilde{\delta}$ into Eq.(3.13) for energy
eigenvalues $E^{(D)}_{n_{r}l}$, we obtain

\begin{equation}
\begin{array}{l} {E^{(D)}_{n_{r}l}=\frac{V_{0}}{2}-\frac{\hbar ^{2}
\tilde{l}(\tilde{l}+1)}{2\mu R_{0}^{2}}\left
(C_{0}+\frac{C_{1}+C_{2}}{2}\right)-\frac{\hbar^{2}}{2\mu
a^{2}}\left[\frac{1}{16}\left(\sqrt{1+\frac{8\mu a^{2}W}{\hbar^{2}
}+\frac{4\tilde{l}(\tilde{l}+1)a^{2}C_{2} }{R_{0}^{2}}}
-2n_{r}-1\right)^{2}+ \right. } \\ \left.{\, \, \, \, \, \, \, \, \,
\, \, \, \frac{\left(\frac{2\mu a^{2}}{\hbar^{2}
}\right)^{2}\left[V_{0}-\frac{\hbar^{2} \tilde{l}(\tilde{l}+1)}{2\mu
R_{0}^{2}}\left(C_{1}+C_{2}
\right)\right]^{2}}{\left(\sqrt{1+\frac{8\mu a^{2}W}{\hbar^{2}
}+\frac{4\tilde{l}(\tilde{l}+1)a^{2}C_{2} }{R_{0}^{2}}}
-2n_{r}-1\right)^{2}}}\right]  \end{array}.
\end{equation}

If the conditions Eqs. (3.14) - (3.16) are satisfied simultaneously,
the bound states will exist. Thus, the energy spectrum Eq. (3.17) is
limited, i.e. we have only the finite number of energy eigenvalues.

For very large $V_{0}$, the $l$-dependent effective potential has
the same form as the potential with $l=0$. According to Eq. (3.17)
the energy eigenvalues depend on the depths of the potential well
$V_{0}$ and $W$, the width $R_{0}$ potential, the thickness $a$
surface, and $D$ parameter. If constraints imposed on $n_{r}$,
$V_{0}$ and  $E_{n_{r}l}^{(D)}$ satisfied, the bound states appear.
Any energy eigenvalue has to satisfy the following inequality:
$V_{eff,min}<E_{n_{r}l}^{(D)}<V_{l}$ , where
$V_{l}=-V_{0}+\frac{\hbar ^{2} \tilde{l}(\tilde{l}+1)}{2\mu
R_{0}^{2}}\left(C_{0}+C_{1}+C_{2} \right)$. From Eq.(3.16) is seen
that the potential depth $V_{0}$ increases when the parameter
$R_{0}$ is decreasing for given $l$ quantum number and vice versa.
Therefore, one can say that the bound states exist within this
potential.

In addition, we have seen that there are some restrictions on the
potential parameters for the bound state solutions within the
framework of quantum mechanics. Hence, when the values of the
parameters $n_{r}$, $V_{0}$ and energy eigenvalues
$E_{n_{r}l}^{(D)}$ satisfy the conditions in Eqs.(3.14) - (3.16) and
$V_{eff,min}<E_{n_{r}l}^{(D)}<V_{l}$ respectively, we obtain the
bound states. We also point out that the exact results obtained for
the generalized Woods-Saxon potential may have some interesting
applications for studying different quantum mechanical and nuclear
scattering problems. Consequently, the found wave functions are
physical ones.

According to the Eq. (3.17), when $l=0$  and $D=3$ , the energy
eigenvalues are obtained the following expression:

\begin{equation}
E_{n_{r}}=\frac{V_{0}}{2}-\frac{\hbar^{2}}{2\mu
a^{2}}\left[\frac{1}{16}\left(\sqrt{1+\frac{8\mu a^{2}W}{\hbar^{2}
}}-2n_{r}-1\right)^{2}+\frac{\left(\frac{2\mu a^{2}V_{0}}{\hbar^{2}
}\right)^{2}}{\left(\sqrt{1+\frac{8\mu a^{2}W}{\hbar^{2}}}
-2n_{r}-1\right)^{2}}\right].
\end{equation}

It should be noted that the Eq. (3.18) is same with the energy
eigenvalues expression in Ref. ~\cite {Editorial}. There the any
energy eigenvalues must be less than $-V_{0}$  and bigger than the
minimum value of generalized Woods-Saxon potential $V_{min}$ , i.e.
$V_{min}<E_{n_{r}}<-V_{0}$, where
$V_{min}=-\frac{(W+V_{0})^{2}}{4W}$.

Now, we are going to define the radial eigenfunctions of this
potential. Having substituted $\pi (z)$ and $\sigma (z)$ into the
equations ~\cite {Nikiforov}

\qquad\qquad\qquad \qquad\qquad $\frac{\Phi '(z)}{\Phi (z)}
=\frac{\pi (z)}{\sigma (z)}$ \qquad  and \qquad $\frac{d}{dz}
(\sigma (z)\rho (z))=\tau (z)\rho (z)$,

one can find the finite function $\Phi (z)$ and weight function
$\rho (z)$ in the interval $(0,\, \, 1)$:

\begin{equation}
\Phi (z)=z^{\varepsilon } (1-z)^{\sqrt{\varepsilon ^{2} -\beta ^{2}
+\gamma ^{2} } } ,
\end{equation}

\begin{equation}
\rho (z)=z^{2\varepsilon } (1-z)^{2\sqrt{\varepsilon ^{2} -\beta
^{2} +\gamma ^{2} } }.
\end{equation}

By substituting the functions $\sigma (z)$ and $\rho (z)$ into
Rodrigues relation ~\cite {Nikiforov} $ y_{n} (z)=\frac{B_{n} }{\rho
(z)} \frac{d^{n} }{dz^{n} } [\sigma ^{n} (z)\rho (z)]$, we get

\begin{equation}
y_{n_{r} } (z)=B_{n_{r} } z^{-2\varepsilon }
(1-z)^{-2\sqrt{\varepsilon ^{2} -\beta ^{2} +\gamma ^{2} } }
\frac{d^{n_{r} } }{dz^{n_{r} } } \left[z^{n_{r} +2\varepsilon }
(1-z)^{n_{r} +2\sqrt{\varepsilon ^{2} -\beta ^{2} +\gamma ^{2} } }
\right],
\end{equation}
where $B_{n_{r}} $ is the normalization constant and its value is
$\frac{1}{n_{r}!} $ ~\cite {Bateman}. After using the following
definition of the Jacobi polynomial

\[P_{n}^{(\alpha \, ,\, \beta )} (1-2z)=\frac{1}{n!} z^{-\alpha } (1-z)^{-\beta } \frac{d^{n} }{dz^{n} }
\left[z^{n+\alpha } (1-z)^{n+\beta } \right],\]

we are able to write Eq.(3.21) as

\[y_{n_{r} } (z)=P_{n_{r} }^{(2\varepsilon \, ,\, \, 2\sqrt{\varepsilon ^{2} -\beta ^{2} +\gamma ^{2} } )}
(1-2z).\]

According to the relation $u(z)=\Phi(z)y(z)$ ~\cite {Nikiforov}, we
obtain the hyper-radial wave functions as:

\begin{equation}
u_{n_{r} l} (z)=C_{n_{r} l} z^{\varepsilon }
(1-z)^{\sqrt{\varepsilon ^{2} -\beta ^{2} +\gamma ^{2} } } P_{n_{r}
}^{(2\varepsilon \, ,\, \, 2\sqrt{\varepsilon ^{2} -\beta ^{2}
+\gamma ^{2} } )} (1-2z).
\end{equation}

Finally, the normalization constant $C_{n_{r}l}$ is determined by
using the following orthogonality condition:

\begin{equation}
\int _{0}^{\infty }\left|R_{n_{r}l} (r)\right|^{2} r^{D-1}  dr=\int
_{0}^{\infty }\left|u_{n_{r}l} (r)\right|^{2} dr=a\int _{0}^{1
}\frac{\left|u_{n_{r}l} (z)\right|^{2} }{z(1-z)}  dz=1 .
\end{equation}

\section{\bf Solution of the hyper-radial Schr\"{o}dinger equation by Supersymmetric quantum mechanics method}

According to Supersymmetric quantum mechanics (SUSY QM), the
eigenfunction of ground state $u_{0} (r)$ in Eq.(2.18) is a form as
below ~\cite {Cooper}

\begin{equation}
u_{0} (r)=N\exp \left(-\frac{\sqrt{2\mu } }{\hbar } \int W(r)dr
\right),
\end{equation}
where $N$ and $W(r)$ are normalized constant and superpotential,
respectively. The connection between the supersymmetric partner
potentials $V_{1}(r)$ and $V_{2} (r)$ of the superpotential $W(r)$
is as follows ~\cite {Cooper}:

\begin{equation}
V_{1} (r)=W^{2} (r)-\frac{\hbar }{\sqrt{2\mu } } W'(r)+E\, \, ,\, \,
\, \, V_{2} (r)=W^{2} (r)+\frac{\hbar }{\sqrt{2\mu } } W'(r)+E .
\end{equation}

The particular solution of the Riccati equation Eq.(4.2) searches
the following form:

\begin{equation}
W(r)=-\frac{\hbar }{\sqrt{2\mu } }
\left(A+\frac{B}{1+e^{\frac{r-R_{0} }{a} } } \right),
\end{equation}
where $A$ and $B$ are unknown constants. Since $V_{1}
(r)=\tilde{V}_{eff} (r)$, inserting the relations Eq.(2.16) and
Eq.(4.3) into the expression Eq.(4.2), and from comparison of
compatible quantities in  the left and right sides of the equation,
we obtain the following relations for determining $A$ and $B$
constants:

\begin{equation}
A^{2} =-\frac{2\mu }{\hbar ^{2}}(E_{0} - \tilde{\delta }C_{0} ) \,
,\, \, \, \, 2AB-\frac{B}{a} =-\frac{2\mu (V_{0}+W-\tilde{\delta
}C_{1})}{\hbar ^{2}} \, ,\, \, \, \, B^{2} +\frac{B}{a}=\frac{2\mu
(W+\tilde{\delta }C_{2})}{\hbar ^{2}} \, .
\end{equation}
So, substituting Eqs.(2.20) into Eqs.(4.4), yields:

\begin{equation}
A^{2} =\frac{\varepsilon ^{2}}{a^{2} } \, ,\, \, \, \,
2AB-\frac{B}{a} =-\frac{\beta ^{2}}{a^{2} } \, ,\, \, \, \, B^{2}
+\frac{B}{a} =\frac{\gamma ^{2}}{a^{2}} \, .
\end{equation}
Having inserted Eq.(4.3) into Eq.(4.1), we find the eigenfunction
for ground state as

\begin{equation}
u_{0} (r)=N\, e^{Ar} \left(1+e^{-\frac{r-R_{0} }{a}} \right)^{-aB} .
\end{equation}
$A$ must be less than zero $(A<0)$, and $B$  must be greater than
zero $(B>0)$ for the hyper-radial $u_{0}(r)$ wave function satisfy
the boundary conditions $u_{0}(0)=0$ and $u_{0} (\infty )=0$. Under
this circumstance, from the second and third equations Eqs.(4.5), we
obtain:

\begin{equation}
A=\frac{1}{2a} -\frac{\beta ^{2} }{a\left(\sqrt{1+4\gamma ^{2} }
-1\right)} \, ,
\end{equation}

\begin{equation}
B=\frac{\sqrt{1+4\gamma ^{2} } -1}{2a} ,
\end{equation}

By using Eq.(4.7) and the first equation of Eq.(4.5), we will obtain
the ground state energy $E_{0}^{(D)}$ as

\begin{equation}
E_{0}^{(D)} =\tilde{\delta }C_{0} -\frac{\hbar ^{2} }{2\mu}
\left[\frac{1}{2a} -\frac{\beta ^{2} }{a\left(\sqrt{1+4\gamma ^{2} }
-1\right)} \right]^{2} .
\end{equation}

When $r\to \infty $, the chosen superpotential $W(\, r)$ is $W(\,
r)\to -\frac{\hbar A}{\sqrt{2\mu }}$. Having inserted the Eq.(4.3)
into Eq.(4.2), for supersymmetric partner potentials, gives:
\begin{equation}
V_{1} (r)=\frac{\hbar ^{2} }{2\mu } \left[A^{2} +\frac{B^{2}
+\frac{B}{a} }{\left(1+e^{\frac{r-R_{0} }{a} } \right)^{2} }
+\frac{2AB-\frac{B}{a} }{1+e^{\frac{r-R_{0} }{a} } } \right]
\end{equation}
and
\begin{equation}
V_{2} (r)=\frac{\hbar ^{2} }{2\mu } \left[A^{2} +\frac{B^{2}
-\frac{B}{a} }{\left(1+e^{\frac{r-R_{0} }{a} } \right)^{2} }
+\frac{2AB+\frac{B}{a} }{1+e^{\frac{r-R_{0} }{a} } } \right] .
\end{equation}

If we add side-by-side the second equation of Eqs.(4.5) to third
equation of Eqs.(4.5), we will obtain:
\begin{equation}
\, 2AB+B^{2} =\frac{\gamma ^{2} -\beta ^{2} }{a^{2} }
\end{equation}
from here
\begin{equation}
\, A=\frac{\gamma ^{2} -\beta ^{2} }{2a^{2} B} -\frac{B}{2} .
\end{equation}

Two partner potentials $V_{1} (\, r)$ and $V_{2} (\, r)$ which
differ from each other with additive constants and have the same
functional form are called invariant potentials
~\cite{Gendenshtein1,Gendenshtein2}. Thus, for the partner
potentials $V_{1} (\, r)$ and $V_{2} (\, r)$ given with Eqs.(4.10)
and (4.11), the invariant forms $R(B_{i} ,  (i=1 , 2 , \ldots ,
n_{r})$, which are independent of $r$, has a form as below:
\begin{equation}
R(B_{1} )=V_{2} (B,\, r)-V_{1} (B_{1} ,r)=-\frac{\hbar ^{2} }{2\mu }
\left[\left(\frac{\gamma ^{2} -\beta ^{2} }{2a^{2}
\left(B-\frac{1}{a} \right)} -\frac{B-\frac{1}{a} }{2} \right)^{2}
-\left(\frac{\gamma ^{2} -\beta ^{2} }{2a^{2} B} -\frac{B}{2}
\right)^{2} \right] ,
\end{equation}
\begin{equation}
\begin{array}{l} R(B_{i} )=V_{2} \left[B-\frac{i-1}{a} \, ,\, r\right]-V_{1}
\left[B-\frac{i}{a} \, ,\, r\right] = \\ -\frac{\hbar ^{2} }{2\mu }
\left[\left(\frac{\gamma ^{2} -\beta ^{2} }{2a^{2}
\left(B-\frac{i}{a} \right)} -\frac{B-\frac{i}{a} }{2} \right)^{2}
-\left(\frac{\gamma ^{2} -\beta ^{2} }{2a^{2} \left(B-\frac{i-1}{a}
\right)} -\frac{B-\frac{i-1}{a} }{2} \right)^{2} \right] .
\end{array}\end{equation}
If we continue this procedure and make the substitution
$\,B_{n_{r}}=B_{n_{r}-1} -\frac{1}{a} =B-\frac{n_{r}}{a}$ at every
step until $\, B_{n_{r}} \ge 0$, the whole discrete spectrum of
Hamiltonian $\, H_{-}(B)$:

\[\begin{array}{l} E_{n_{r}l}^{(D)} =E_{0}^{(D)} +\sum
_{i=1}^{n_{r}}R(B_{i} )= \tilde{\delta }C_{0} -\frac{\hbar ^{2}
}{2\mu } \left[\left(\frac{\gamma ^{2} -\beta ^{2} }{2a^{2}
\left(B-\frac{n_{r}}{a} \right)} -\frac{B-\frac{n_{r}}{a} }{2}
\right)^{2} -\left(\frac{\gamma ^{2} -\beta ^{2} }{2a^{2}
\left(B-\frac{n_{r}-1}{a} \right)} -\frac{B-\frac{n_{r}-1}{a} }{2}
\right)^{2} + \right. \\  \left(\frac{\gamma ^{2} -\beta ^{2}
}{2a^{2} \left(B-\frac{n_{r}-1}{a} \right)}
-\frac{B-\frac{n_{r}-1}{a} }{2} \right)^{2} - {\, \,
\left(\frac{\gamma ^{2} -\beta ^{2} }{2a^{2}
\left(B-\frac{n_{r}-2}{a} \right)} -\frac{B-\frac{n_{r}-2}{a} }{2}
\right)^{2} + \ldots + \left(\frac{\gamma ^{2} -\beta ^{2} }{2a^{2}
\left(B-\frac{2}{a} \right)} -\frac{B-\frac{2}{a} }{2} \right)^{2}-}
\\ {\, \, \left.\left(\frac{\gamma ^{2} -\beta ^{2} }{2a^{2} \left(B-\frac{1}{a}
\right)} -\frac{B-\frac{1}{a} }{2} \right)^{2} + \left(\frac{\gamma
^{2} -\beta ^{2} }{2a^{2} \left(B-\frac{1}{a} \right)}
-\frac{B-\frac{1}{a} }{2} \right)^{2} - \left(\frac{\gamma ^{2}
-\beta ^{2} }{2a^{2} B} -\frac{B}{2} \right)^{2} +\left(\frac{1}{2a}
-\frac{\beta ^{2} }{a\left(\sqrt{1+4\gamma ^{2} } -1\right)}
\right)^{2} \right]} \\  =\tilde{\delta }C_{0} -\frac{\hbar ^{2}
}{2\mu } \left[\frac{\gamma ^{2} -\beta ^{2} }{2a^{2}
\left(B-\frac{n_{r}}{a} \right)} -\frac{B-\frac{n_{r}}{a} }{2}
\right]^{2}= \tilde{\delta }C_{0} -\frac{\hbar ^{2} }{2\mu \, a^{2}
} \left[\frac{\beta ^{2} -\gamma ^{2} }{\sqrt{1+4\gamma ^{2} }
-2n_{r}-1} +\frac{\sqrt{1+4\gamma ^{2} } -2n_{r}-1}{4} \right]^{2} ,
\end{array}\]
and we obtain:
\begin{equation}
E_{n_{r}l}^{(D)} =\tilde{\delta }C_{0} -\frac{\hbar ^{2} }{2\mu \,
a^{2} } \left[\frac{\beta ^{2} -\gamma ^{2} }{\sqrt{1+4\gamma ^{2}}
- 2n_{r}-1} +\frac{\sqrt{1+4\gamma ^{2} } -2n_{r}-1}{4} \right]^{2}
\end{equation}

Thereby, if the parameters $\beta \,,\,\gamma \, ,
\widetilde{\delta}$ are considered into Eq.(4.16), the obtained
expression for energy eigenvalue in $l$-state will be same with
expression Eq.(3.17) which was obtained by NU method. It should be
noted that the same conditions for $n_{r}$ and $V_{0}$ in
Eqs.(3.14)-(3.16) obtained by NU method are also determined from the
following inequalities $B>0, \, A<0$. Thus, the determined
conditions for $n_{r}$, $V_{0}$ and $E_{n_{r}l}^{D)}$, i.e., if the
conditions the Eqs.(3.14)-(3.16) and
$V_{eff,min}<E_{n_{r}l}^{D)}<V_{l}$ are satisfied respectively,
there are the bound states in the system, and the energy spectrum of
these states is limited number. According to the SUSY QM metod,
knowing the ground state eigenvalues $E_{0}^{(D)}$ and
eigenfunctions $u_{0}(r)$, all energy eigenvalues $E_{n_{r}l}^{(D)}$
and eigenfunctions $u_{n_{r}l}(r)$ will be readily obtain. So, the
obtained solution of hyper - radial Schr\"{o}dinger equation by
using the eigenfunction of the ground state is exactly equal with
the solution obtained by using NU method.

\section{\bf Results and Discussion}

In this chapter, in order to analyze the present qualitative
findings, the single particle energy levels, the effective
potentials and normalized hyper - radial wave functions of neutron
moving under the average potential field of the $^{56} Fe$ nucleus
in the form the generalized Woods-Saxon potential field, are
calculated for various values of $n_{r}$ and $l$ quantum numbers by
using the empirical values $r_{0}=1.285\, \, fm$ and $a=0.65\, \,
fm$ taken from Ref.~\cite {Perey}. Under these certain
circumstances, the potential depth of mentioned potential is $V_{0}
=(40.5+0.13A)\, MeV=47.78\, \, MeV$, and the radius of the nucleus
is $R_{0} =r_{0} A^{{\tfrac{1}{3}} } =4.9162\, \, fm$. Here $A$ is
the atomic mass number of $^{56} Fe$ nucleus. The reduced mass
consists of neutron mass $m_{n}=1.00866\, \, u$ and $^{56} Fe$ core
mass with is $m_{A}=56\, \, u$, and its value is $\mu =\frac{m_{A}
\cdot m_{n}}{m_{A} +m_{n}}=0.990814\, \, u$.

It would be noted that the dependent term on the orbital quantum
number $l$ is as $\widetilde{l} (\widetilde{l}+1)$  in the energy
spectrum expresion $E_{n_{r}l}^{(D)}$  the Eq (3.17), and
$\widetilde{l} (\widetilde{l}+1)$ is invariant for given $N$
positive numbers because of the conversion $l'=l \pm N$ and $D'=D
\mp 2N$ , i.e.,$\widetilde{l'} (\widetilde{l'}+1)=\widetilde{l}
(\widetilde{l}+1)$ based on the following conversion:
$\widetilde{l'}= l'+\frac{D'-3}{2}=l \pm N + \frac{D \mp 2N-3}{2}=l
\pm N+\frac{D-3}{2} \mp N=l+\frac{D-3}{2}=\widetilde{l}$. Therefore,
when $(n_{r}, l, D) \longrightarrow (n_{r}, l \pm N, D \mp 2N)$, the
others energy spectrum in high $D$ dimension can be found as
$E_{n_{r}l}^{(D)}=E_{n_{r};\; l \pm N}^{(D \mp 2N)}$ with knowing
the energy spectrum in certain dimensions. When $D\geq 2, N$ is
equal 1. Hence, it is sufficient to find the energy spectrum in
$D=2$ and $D=3$, because the energy spectrum in high $D$ dimension
can be determined by using the following relation:\qquad
$E_{n_{r}l}^{(D)}=E_{n_{r};\; l \pm 1}^{(D \mp 2)}$, i.e.,
\begin{equation}
E_{n_{r};\; l}^{(4)}=E_{n_{r};\;l+1}^{(2)} \;,  \; \;  E_{n_{r};\;
l}^{(6)}=E_{n_{r};\;l+2}^{(2)} \;,  \; \; E_{n_{r};\;
l}^{(8)}=E_{n_{r};\;l+3}^{(2)} \;,\; \ldots
\end{equation}
and
\begin{equation}
E_{n_{r};\; l}^{(5)}=E_{n_{r};\;l+1}^{(3)} \;,  \; \;  E_{n_{r};\;
l}^{(7)}=E_{n_{r};\;l+2}^{(3)} \;,  \; \; E_{n_{r};\;
l}^{(9)}=E_{n_{r};\;l+3}^{(3)} \;,\; \ldots
\end{equation}

In the case of $D=2$  and $D=3$ , the minimum point $r_{min}$  of
effective generalized Woods-Saxon potential and its minimum value
$V_{eff, min}$ , the value  $V_{l}=-V_{0}+\frac{\hbar ^{2}
\tilde{l}(\tilde{l}+1)}{2\mu R_{0}^{2}}\left(C_{0}+C_{1}+C_{2}
\right)$, the bound state energies and the corresponding normalized
hyper-radial wave functions are given for certain numbers $l$ and
$n_{r}$ in the Table 1 and Table 2, respectively. Additionally, when
$D=2$ and $D=3$, the normalized hyper-radial wave functions of the
bound state as a function of $z$ are plotted for certain numbers $l$
and $n_{r}$ at $W=100\; MeV$ in Fig. 1 and Fig. 2, respectively.

It seen from the Table 1 and Table 2 that when $W=100 \;MeV$  and
$W=200 \;MeV$, there are respectively bound states for $n_{r}=0$ in
$0\leq l \leq 5$ and $0\leq l \leq 6$, but there are not
respectively bound states in $l \geq 6$ and $l \geq 7$  as well as
when $n_{r} \geq 1$, there are not bound states in any $l$.
Furthermore, it seen also from the Table 1 and Table 2 that while
the value of $W$ increases, the minimum point of potential $r_{e}$
gets closer to $R_{0}$. Besides, while the orbital quantum number
$l$ increases, the energy of bound states $E_{n_{r}l}^{(D)}$ also
increase in the setted same value of the radial quantum number
$n_{r}$ and $D$ parameter as well as the increasing of $D$ dimension
leds to the increasing of bound state energies,
i.e.,$E_{01}^{(3)}>E_{0l}^{(2)}\;, \; E_{02}^{(3)}>E_{02}^{(2)}\;,
\;E_{03}^{(3)}>E_{03}^{(2)}\;$ .

It means that, in both cases, there is occurring the pushing force
at the expense of the additional centrifugal potential as
$V_{l}(r)=\frac{\hbar ^{2} \tilde{l}(\tilde{l}+1)}{2\mu r^{2}}$.
Therefore, both the number and value of the energy bound state must
increase in order to compensate for this potential \cite{Greiner,
Ballentine}. It ought to be noted that the inequality (3.14) and
(3.16) are satisfied, but the inequality (3.15) are not satisfied
for $n_{r}=1$  and $1 \leq l \leq 5$ case in $W=100 \;MeV$ value in
Table 1. It means any of these findings for $n_{r}=1$  and $1 \leq l
\leq 5$ cannot be considered as a physical and are only the
mathematical results. It has also been determined that when $W=200
\;MeV$, there are not bound states for $n_{r}=1$  and $0 \leq l \leq
6$. Hence, in $W=100 \;MeV$,  and $W=200 \;MeV$, the non-obtaining
bound states of hyper-radial Schr\"{o}dinger equation for the
generalized Woods-Saxon potential in $n_{r}=1$  and $1 \leq l \leq
5$, and $n_{r}=1$  and $0 \leq l \leq 6$, cases can only be
eliminated by using the spin and pseudospin symmetry
~\cite{Hecht,Arima,Ginocchio1,Zhou,Ginocchio2,Liang} in this type
potentials.

\section{\bf Conclusion}

To conclude, an analytical study of the hyper-radial Schr\"{o}dinger
equation have been performed for the generalized Woods-Saxon
potential by using the improved approximation scheme to the
centrifugal term for any $l$ -states. There the energy eigenvalues
of the bound states and corresponding eigenfunctions have been
analytically found via both of NU and SUSY QM methods within the
Pekeris approximation. The same expressions were obtained for the
energy eigenvalues as well as the expression of hyper-radial wave
functions transformed each other was shown by using mentioned
methods. The energy eigenvalues depending on $V_{0}\, ,\,W \, , \,
R_{0} \, ,\, a$ and $D$ parameters have a finite number energy
spectrum for the generalized Wood-Saxon potential, so it puts some
limitation on the potential parameters during the solution of the
bound states within the framework of quantum mechanics. In this way,
if the parameters $V_{0}$, $n_{r}$   and energy eigenvalues
$E_{n_{r}l}^{(D)}$ satisfy the conditions in Eqs.(3.14) - (3.16) and
$V_{eff,min}<E_{n_{r}l}^{(D)}<V_{l}$  respectively, it means there
are the bound states in system. It ought to be noted that the same
limiting conditions were obtained for  $V_{0}$ and $n_{r}$  thanks
to both methods. Since there is the practical interest for the
energy spectrum in various potential, investigating the features of
eigenvalues is very important and actual with regard to arbitrary
parameter of system. For illustration, the bound state energies and
corresponding normalized hyper-radial wave functions of neutron
moving under the generalized Woods-Saxon potential field, have been
calculated and analyzed for some  $l$ and $n_{r}$ values.

\newpage
\begin{table}[h]
\parindent=-1.5cm
\begin{tabular}{|c|c|c|c|c|c|c|c|}\hline
$n_{r} $ & $l$ & $W, MeV  $ & $r_{min} ,\;fm $ & $V_{eff min} , MeV$ & $E_{n_{r}l}^{(D)} , MeV$ & $V_{l} , MeV$ & $u_{n_{r}l}(z) $ \\
\hline

0 & 0 & 100 & 4.240031608 & -54.89115601 & -48.28466604 &
-48.27312422 & $0.2193602109z^{0.9767738828}(1-z)^{0.01519140062} $
\\ \hline

0 & 1 & 100 & 4.240031608 & -53.71581608 & -46.38731403 &
-46.30062733 & $0.3721331375z^{0.9806163276}(1-z)^{0.04163292552} $
\\ \hline

1 & 1 & 100 & 4.240031608 & -53.71581608 & -23068.47035 &
-46.30062733 & $Unbound$
\\ \hline

0 & 2 & 100 & 4.240031608 & -50.18979628 & -41.03113420 &
-40.38313667 & $0.6597273740z^{1.045547623}(1-z)^{0.1138275476} $ \\
\hline

1 & 2 & 100 & 4.240031608 & -50.18979628 & -1016.893447 & -40.38313667 & $Unbound$ \\
\hline

0 & 3 & 100 & 4.240031608 & -44.31309661 & -32.88286462 &
-30.52065223 & $0.9860130431z^{1.027256859}(1-z)^{0.2173302481} $ \\
\hline

1 & 3 & 100 & 4.240031608 & -44.31309661 & -218.3838905 & -30.52065223 & $Unbound$ \\
\hline

0 & 4 & 100 & 4.240031608 & -36.08571707 & -22.45213564 &
-16.71317402 & $1.359442712z^{1.079307764}(1-z)^{0.3387484872} $ \\
\hline

1 & 4 & 100 & 4.240031608 & -36.08571707 & -71.52563286 & -16.71317402 & $Unbound$ \\
\hline

0 & 5 & 100 & 4.240031608 & -25.50765766 & -9.98007814 &
1.03929797 & $1.788974091z^{1.150936539}(1-z)^{0.4693958088} $ \\
\hline

1 & 5 & 100 & 4.240031608 & -25.50765766 & -15.77873438 & 1.03929797 & $Unbound$ \\
\hline

0 & 6 & 100 & 4.240031608 & -12.57891838 & 4.45561004 & 22.73676374 & $Unbound$ \\
\hline

0 & 0 & 200 & 4.599510648 & -76.99336041 & -59.34650380 &
-48.20458910 &  $1.747348676z^{1.085457023}(1-z)^{0.4719985004} $ \\
\hline

1 & 0 & 200 & 4.599510648 & -76.99336041 & -64.91459392 &
-48.20458910 & $Unbound$
\\ \hline

0 & 1 & 200 & 4.599510648 & -75.99456066 & -58.16160295 &
-46.50623270 & $1.781805687z^{1.090041485}(1-z)^{0.4827516215} $ \\
\hline

1 & 1 & 200 & 4.599510648 & -75.99456066 & -61.50356910 &
-46.50623270 & $Unbound$
\\ \hline

0 & 2 & 200 & 4.599510648 & -72.99816142 & -53.74780856 &
-41.41116348 & $1.826628107z^{1.095871056}(1-z)^{0.4966600632} $ \\
\hline

1 & 2 & 200 & 4.599510648 & -72.99816142 & -51.14073127 &
-41.41116348 & $Unbound$
\\ \hline

0 & 3 & 200 & 4.599510648 & -68.00416269 & -46.81924259 &
-32.91938145 & $1.927039632z^{1.109380285}(1-z)^{0.5271885085} $ \\
\hline

1 & 3 & 200 & 4.599510648 & -68.00416269 & -36.27696168 &
-32.91938145 & $Unbound$
\\ \hline

0 & 4 & 200 & 4.599510648 & -61.01256446 & -37.35316172 &
-21.03088661 & $2.077111816z^{1.130092659}(1-z)^{0.5712826715} $ \\
\hline

1 & 4 & 200 & 4.599510648 & -61.01256446 & -18.70310516 &
-21.03088661 & $Unbound$
\\ \hline

0 & 5 & 200 & 4.599510648 & -52.02336675 & -25.51726932 &
-5.74567896 & $2.282942428z^{1.159069026}(1-z)^{0.6287550789} $ \\
\hline

1 & 5 & 200 & 4.599510648 & -52.02336675 & 0.19801712 & -5.74567896
& $Unbound$
\\ \hline

0 & 6 & 200 & 4.599510648 & -41.03656954 & -10.78664839 &
12.93624150 & $2.511668721z^{1.191311742}(1-z)^{0.6887228640} $ \\
\hline

1 & 6 & 200 & 4.599510648 & -41.03656954 & 22.67135873 & 12.93624150
& $Unbound$
\\ \hline

0 & 7 & 200 & 4.599510648 & -28.05217283 & 4.69258483 & 35.01487477
& $Unbound$
\\ \hline

\end{tabular}

\caption{Calculated the bound state energies and corresponding
normalized hyper-radial wave functions for $V_{0} =47.78 MeV$,
$R_{0} =4.9162 fm$, $a=0.65 fm$, $D=2$ in different values of
$n_{r}$ and $l$.} \label{table1}
\end{table}

\newpage
\begin{table}[h]
\parindent=-1.5cm
\begin{tabular}{|c|c|c|c|c|c|c|c|}\hline
$n_{r} $ & $l$ & $W, MeV  $ & $r_{min} ,\;fm $ & $V_{eff min} , MeV$ & $E_{n_{r}l}^{(D)} , MeV$ & $V_{l} , MeV$ & $u_{n_{r}l}(z) $ \\
\hline

0 & 0 & 100 & 4.240031608 & -54.59732103 & -47.80412135 & -47.8 &
$0.2654290888z^{0.9776724433}(1-z)^{0.02196147521} $
\\ \hline

0 & 1 & 100 & 4.240031608 & -52.24664116 & -44.10044208 &
-43.83500662 & $0.5061434438z^{0.9862584723}(1-z)^{0.07285183540} $
\\ \hline

1 & 1 & 100 & 4.240031608 & -52.24664116 & -3361.266693 &
-43.83500662 & $Unbound$
\\ \hline

0 & 2 & 100 & 4.240031608 & -47.54528142 & -37.26729879 &
-35.94501867 & $0.8156347139z^{1.008916222}(1-z)^{0.1626007103} $ \\
\hline

1 & 2 & 100 & 4.240031608 & -47.54528142 & -429.1546502 &
-35.94501867 & $Unbound$
\\ \hline

0 & 3 & 100 & 4.240031608 & -40.49324182 & -27.93239468 &
-24.1100373 & $1.166824188z^{1.050720153}(1-z)^{0.2764562078} $ \\
\hline

1 & 3 & 100 & 4.240031608 & -40.49324182 & -123.0701748 &
-24.1100373 & $Unbound$
\\ \hline

0 & 4 & 100 & 4.240031608 & -31.09052234 & -16.46401658 &
-8.33006225 & $1.455628707z^{0.9352553768}(1-z)^{0.40328453581} $ \\
\hline

1 & 4 & 100 & 4.240031608 & -31.09052234 & -39.07468690 &
-8.33006225 & $Unbound$
\\ \hline

0 & 5 & 100 & 4.240031608 & -19.33712300 & -3.00613603 &
11.39400663 & $2.031457706z^{1.193259909}(1-z)^{0.5366086544} $ \\
\hline

1 & 5 & 100 & 4.240031608 & -19.33712300 & 2.85638509 & 11.39400663
& $Unbound$
\\ \hline

0 & 6 & 100 & 4.240031608 & -5.23304379 & 12.40489010 & 35.06486928
& $Unbound$
\\ \hline

0 & 0 & 200 & 4.599510648 & -76.74366047  & -59.04960487 & -47.8 &
$1.755963468z^{1.086598753}(1-z)^{0.4746954280} $ \\ \hline

1 & 0 & 200 & 4.599510648 & -76.74366047  & -64.04871484 & -47.8 &
$Unbound$
\\ \hline

0 & 1 & 200 & 4.599510648 & -74.74606098 & -56.21001797 &
-44.38328719 & $1.793100893z^{1.091444217}(1-z)^{0.4862874478} $ \\
\hline

1 & 1 & 200 & 4.599510648 & -74.74606098 & -56.94732674 &
-44.38328719 & $Unbound$
\\ \hline

0 & 2 & 200 & 4.599510648 & -70.7508619 & -50.61117817 &
-37.58986157 & $1.871017637z^{1.101800350}(1-z)^{0.5102560391} $ \\
\hline

1 & 2 & 200 & 4.599510648 & -70.7508619 & -44.16267043 &
-37.58986157 & $Unbound$
\\ \hline

0 & 3 & 200 & 4.599510648 & -64.75806351 & -42.39278162 &
-27.39972313 & $1.995480479z^{1.118763012}(1-z)^{0.5475273550} $ \\
\hline

1 & 3 & 200 & 4.599510648 & -64.75806351 & -27.72415839 &
-27.39972313 & $Unbound$
\\ \hline

0 & 4 & 200 & 4.599510648 & -56.76766554 & -31.72137742 &
-13.81287189 & $2.032995412z^{1.018948454}(1-z)^{0.5983983356} $ \\
\hline

1 & 4 & 200 & 4.599510648 & -56.76766554 & -9.36267859 &
-13.81287189 & $Unbound$
\\ \hline

0 & 5 & 200 & 4.599510648 & -46.77966808 & -18.75894891 &
3.17069217 & $2.408652680z^{1.176882622}(1-z)^{0.6621806199} $ \\
\hline

1 & 5 & 200 & 4.599510648 & -46.77966808 &  9.92379481 & 3.17069217
& $Unbound$
\\ \hline

0 & 6 & 200 & 4.599510648 & -34.79407112 & -3.64144188 &
23.55096903 & $2.709803436z^{1.219329800}(1-z)^{0.7373684670} $ \\
\hline

1 & 6 & 200 & 4.599510648 & -34.79407112 &  29.81656524 &
23.55096903 & $Unbound$
\\ \hline

0 & 7 & 200 & 4.599510648 & -20.81087467 &  13.53043692 &
46.32795871 & $Unbound$
\\ \hline

\end{tabular}

\caption{Calculated the bound state energies and corresponding
normalized hyper-radial wave functions for $V_{0} =47.78 MeV$,
$R_{0} =4.9162 fm$, $a=0.65 fm$, $D=3$ in different values of
$n_{r}$ and $l$.} \label{table1}
\end{table}

\newpage


\begin{figure}
\vskip 1.2cm\epsfxsize 9.8cm \centerline{\epsfbox{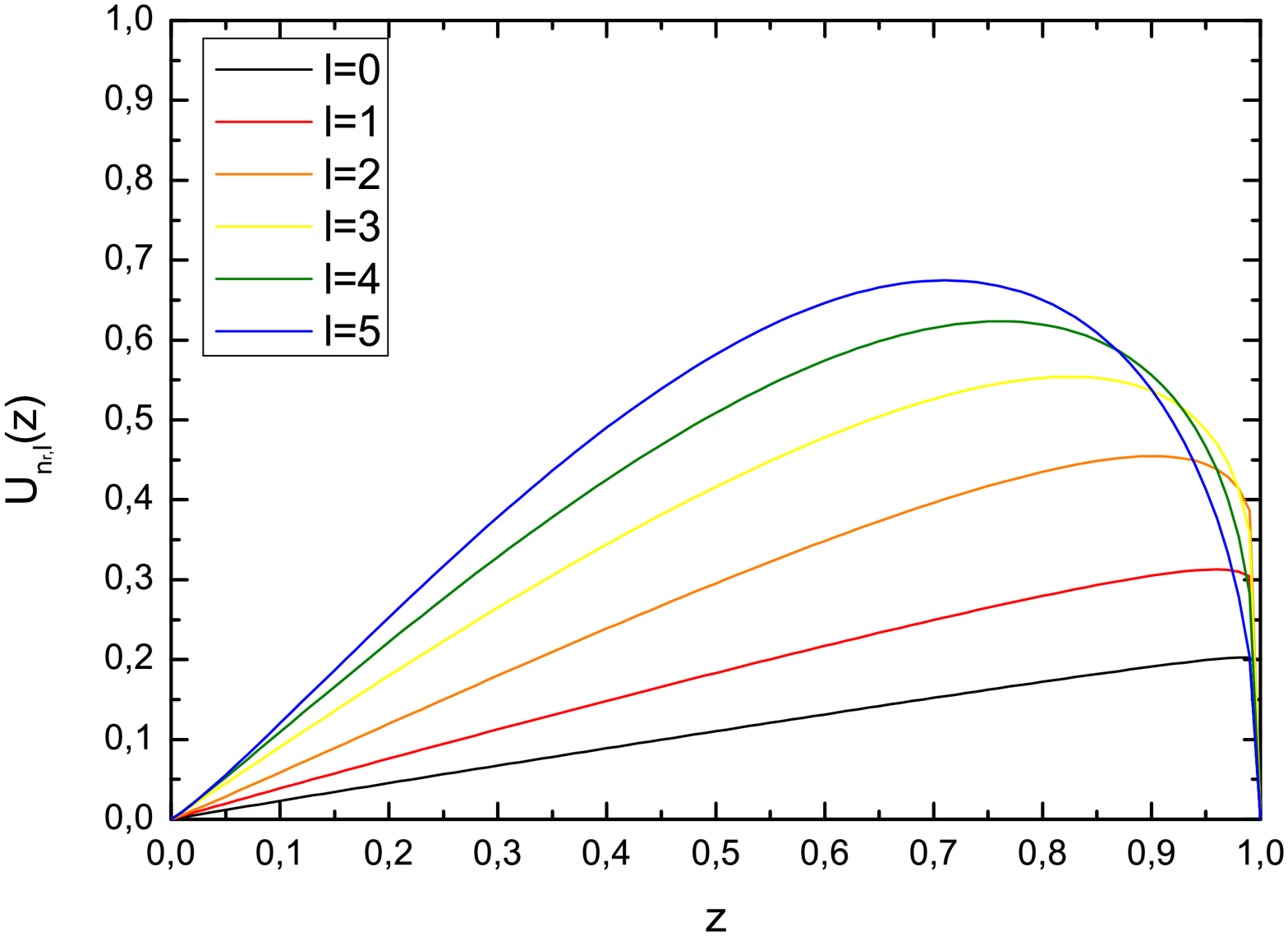}}

\vskip-0.2cm \caption{The normalized hyper-radial wave functions
$u_{n_{r}l}(z)$ as a function of $z$ and several quantum numbers
$l$ for $V_{0} =47.78 MeV$, $R_{0} =4.9162 fm$, $a=0.65 fm$,
$D=2$, $n_{r}=0$.} \label{Fig1}
\end{figure}

\begin{figure}
\vskip 1.2cm\epsfxsize 9.8cm \centerline { \epsfbox{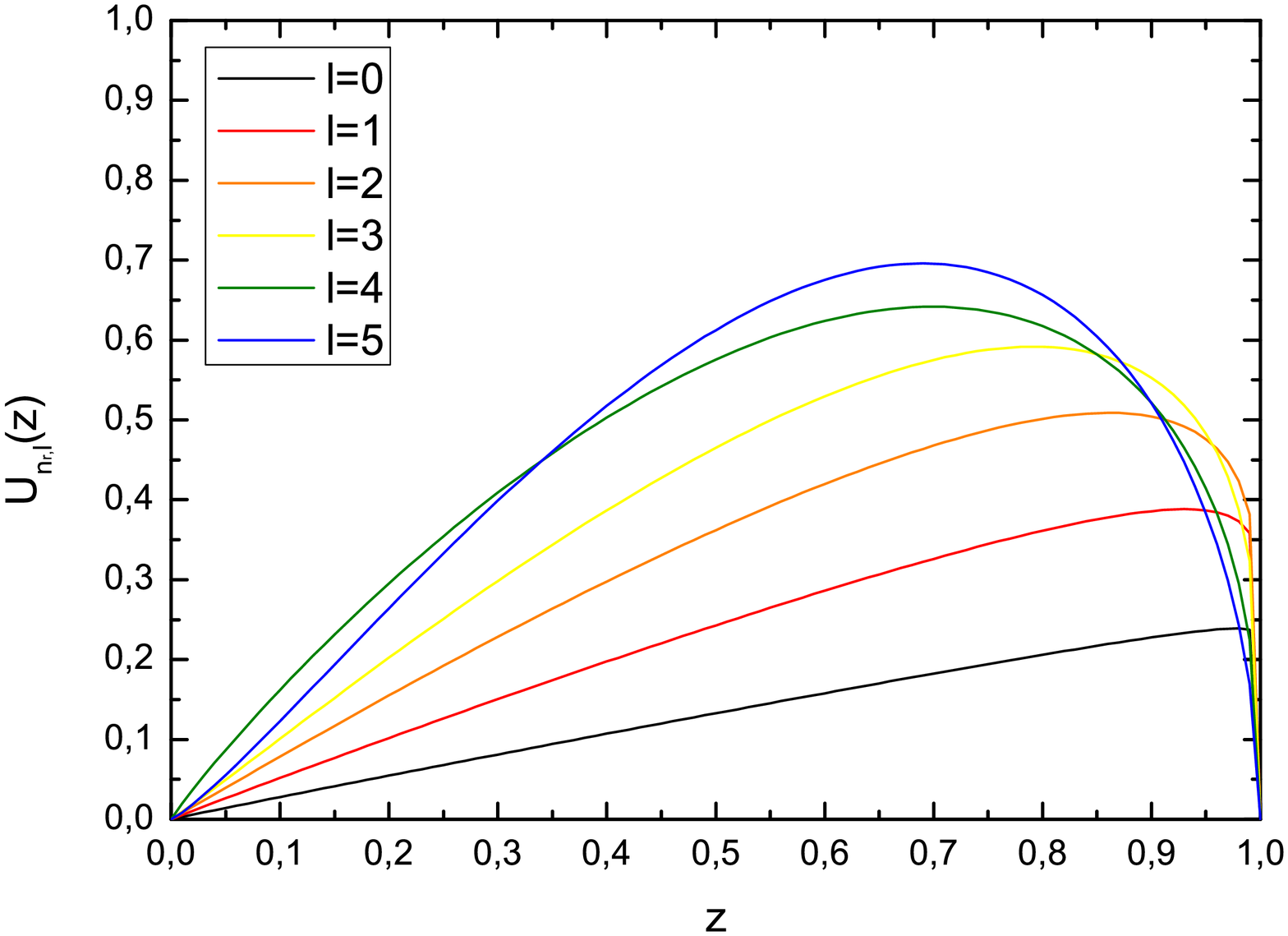}}

\vskip-0.2cm \caption{The normalized hyper-radial wave functions
$u_{n_{r}l}(z)$ as a function of $z$ and several quantum numbers
$l$ for $V_{0} =47.78 MeV$, $R_{0} =4.9162 fm$, $a=0.65 fm$,
$D=3$, $n_{r}=0$.} \label{Fig2}
\end{figure}

\newpage

\end{document}